# Accelerating vapor condensation with daytime radiative cooling

Ming Zhou[1], Haomin Song[2], Xingyu Xu[3], Alireza Shahsafi[1], Zhenyang Xia[1], Zhenqiang Ma[1], Mikhail A. Kats[1], Jia Zhu[4], Boon S. Ooi[5], Qiaoqiang Gan[2, †], Zongfu Yu[1, *]

1. Department of Electrical and Computer Engineering, University of Wisconsin – Madison, Madison, WI 53705, USA
2. Department of Electrical Engineering, The State University of New York at Buffalo, Buffalo, NY 14260, USA
3. School of Materials Science and Engineering, Tsinghua University, Beijing 100084, China
4. College of Engineering and Applied Sciences, Nanjing University, Nanjing 210093, China
5. Photonics Laboratory, King Abdullah University of Science and Technology (KAUST), Thuwal, 23955-6900, Saudi Arabia

† Correspondence to: Qiaoqiang Gan (qqgan@buffalo.edu)

* Correspondence to: Zongfu Yu (zyu54@wisc.edu)

**Vapor condensation plays a crucial role in solar water-purification technologies. Conventional condensers in solar water-purification systems do not provide sufficient cooling power for vapor condensation, limiting the water production rate to 0.4 L m$^{-2}$ hour$^{-1}$. On the other hand, radiative dew condensation, a technique used by existing radiative dew condensers, only works at nighttime and is incompatible with solar water-purification technologies. Here, we develop daytime radiative condensers that reflect almost all solar radiation, and can thus create dew water even in direct sunlight. Compared to state-of-art condensers, our daytime radiative condenser doubles the production of purified water over a 24-hour period. The integration of our daytime radiative condenser with solar water-purification systems can increase the water production rate in sunlight from 0.4 L m$^{-2}$ hour$^{-1}$ to more than 1 L m$^{-2}$ hour$^{-1}$.**

# Introduction

Energy and clean water are global challenges that are intertwined in an unfavorable way: even in areas where water is available, energy may not be available to purify it for human use[1,2]. In this context, there has been strong interest in developing passive water-purification technology that does not actively consume energy. Solar water-purification technology is particularly promising because solar energy is abundant and widely accessible.

Every solar water-purification system comprises a solar absorber and a condenser. The solar absorber evaporates surface water[3–7] (e.g. seawater) or water vapor adsorbed in a porous metal-organic framework[8], which then condenses on the condenser, which dissipates heat into the surrounding environment. Much progress has been made to reduce the cost of solar absorbers[9] and to maximize the evaporation efficiency by localizing heat to nanoscale volumes[10–13], resulting in evaporation efficiencies approaching the theoretical limit[11] of 1.6 L m$^{-2}$ hour$^{-1}$, which assumes that all the heating power of unconcentrated sunlight[10] (1000 W m$^{-2}$) can be utilized for evaporation.

However, this evaporated water must also be efficiently condensed. Existing solar water-purification systems use passive convection and conduction as the cooling mechanisms for condensation[14–17], and the resulting cooling power is much lower than the heating power of sunlight. The difference results in mismatched evaporation and condensation rates, which pins the water production rate to 0.4 L m$^{-2}$ hour$^{-1}$, a quarter of the theoretical limit of evaporation[18]. There has not been much progress on overcoming this imbalance, which limits the usage of solar water-purification systems in practice.

A different passive water-generation technique is radiative dew condensation. This technique can be found in nature, for example in darkling beetles in the Namib desert[19]. The Namib beetle's body functions as a cooler by shedding thermal energy through mid-infrared (mid-IR) radiation toward a clear nighttime sky, generating dew from humid air. This mechanism is also used by commercial radiative dew condensers[20–22]. Unfortunately, existing radiative dew condensers are fundamentally incompatible with solar water-purification systems since they only work at nighttime, because heating from sunlight overwhelms the radiative cooling power. To integrate a radiative dew condenser with a solar water-purification system, solar absorption must be minimized[20]. However, this goal has never been achieved, as even the best radiative dew condensers absorb more than 15% of sunlight, making them evaporators during the daytime[20].

Recently, Fan et al. showed that passive radiative cooling to sub-ambient temperatures can be realized even during the daytime, by integrating a high-efficiency solar reflector with a high-emissivity thermal emitter in the mid-IR atmospheric transparency window[23]. Using this work as a basis, here we demonstrate a daytime radiative condenser. Compared to existing radiative dew condensers[20–22], our condenser can function even in the presence of sunlight, which is essential for integration into solar water-purification systems that mainly operate during daytime. Consequently, our condenser provides condensation rates that can match the solar evaporation rate, leading to water production rate of more than 1 L m$^{-2}$ hour$^{-1}$.

## Results

Water vapor condenses when its temperature drops below the dew point, which is lower than the ambient temperature[24] when the relative humidity is below 100%. In such an environment, the only passive cooling mechanism that can enable condensation is thermal radiation, because conduction and convection push the condenser temperature toward the ambient temperature. For example, the black body of the Namib beetles strongly emits infrared radiation around the atmospheric-transparency window[19] (8 ~ 13 µm). Only radiation from the universe emits back to the beetle in this band, of which there is little because the background temperature of the universe is about 3 K [25]. Consequently, the Namib beetle can passively cool itself to below the dew point[19].

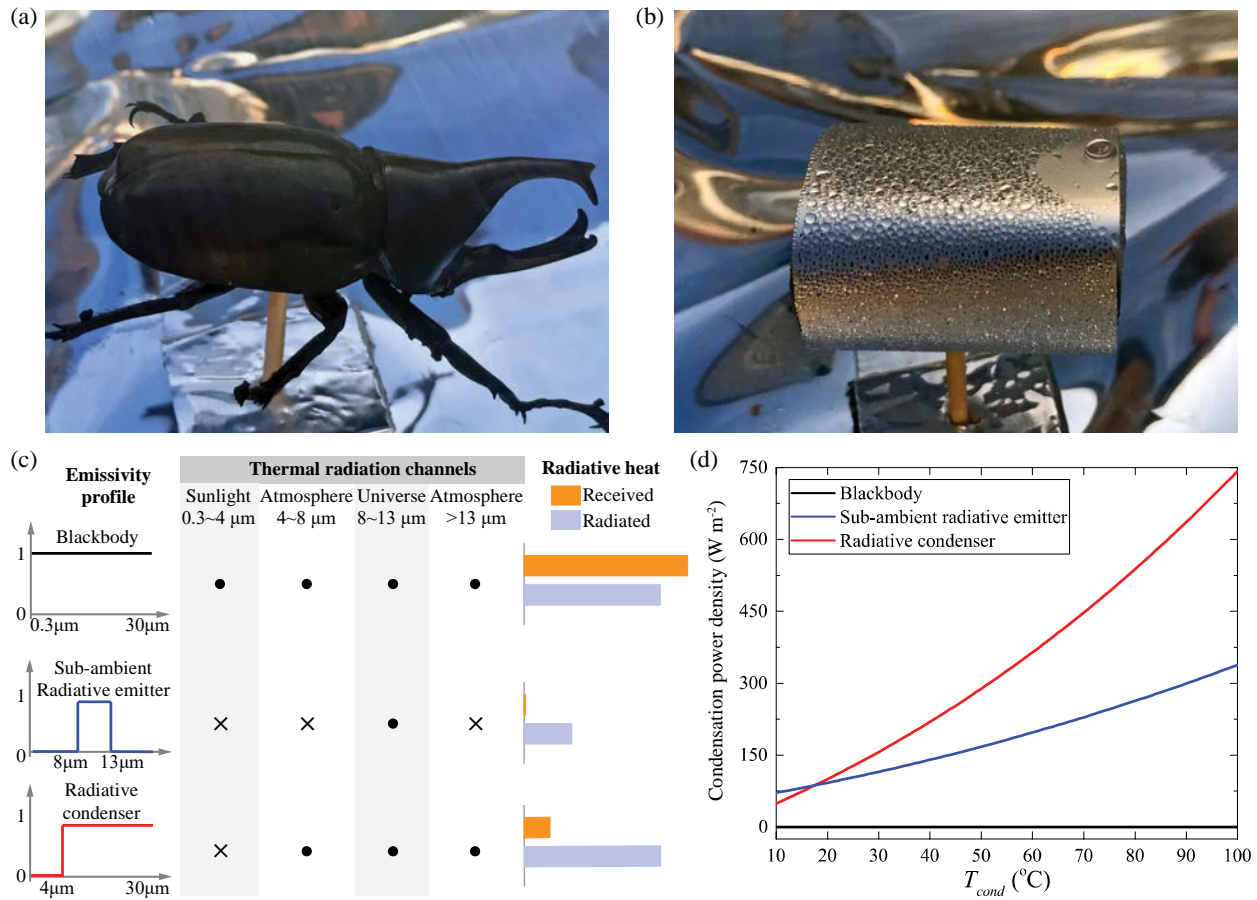

Figure 1. Photos of (a) a dead black beetle and (b) our daytime-condensing device, both placed under sunlight on the roof of a parking ramp at the University of Wisconsin-Madison. We introduce humid air flow across both, with 90% ~ 95% humidity. The experiment was performed on April 26th, 2018. Unlike the black beetle, our device reflects most of the sunlight, and thus condenses water vapor during daytime. (c) Thermal radiation channels and emissivity spectra of a blackbody (black), the sub-ambient radiative emitter (blue) from Ref. [23], and our radiative condenser (red). The spectrum of the radiation is divided into four channels: the solar channel (0.3 ~ 4 µm), the universe channel (8 ~ 13 µm) and the two atmospheric channels (4 ~ 8 µm and >13 µm). The dot (cross) indicates that the emitter has the corresponding channel open (closed) for radiative exchange. The radiative heat received and radiated by the emitters at 100 °C are plotted as orange and blue bars, respectively. (d) Calculated condensation power of the emitters in direct sunlight, operating at different temperatures. The ambient temperature is fixed at 20 °C. The blackbody (black) has zero condensation power. The radiative condenser (red) has much more condensation power than the sub-ambient radiative emitter (blue).

Both the Namib beetle and existing radiative dew condensers only work before the sun is at full strength. As the sun rises, the heat absorbed by a blackbody from solar radiation can reach 1000 W m$^{-2}$ [11], much stronger than the radiation power density of a blackbody at $T_{BB} = 20\ °C$, which is $\sigma T_{BB}^4 \approx 420$ W m$^{-2}$, where $\sigma$ is the Stefan-Boltzmann constant. As a result, no condensation can be achieved at daytime. To enable daytime condensation in sunlight, we spectrally engineer the absorptivity such that the absorption of sunlight in the visible and near-infrared regions is minimized[20,23], enabling condensation during both the day and night. Figure 1a-b shows a black beetle and our device—a solar-blocking infrared-emitting device under direct sunlight. The black beetle absorbs most of the sunlight, suppressing condensation, while our device (Fig. 1b) condenses a considerable amount of water on its surface.

Now we explain the design principle of this radiative condenser. The key is to spectrally engineer the radiative surface to enable or disable specific heat-exchange channels with the environment. The spectrum of the radiation can be roughly divided into four segments under a clear sky during the daytime, as shown in Fig. 1c. In the wavelength range of 0.3 to 4 µm, the incoming radiation is dominated by solar radiation. From 4 to 8 µm, the incoming radiation is dominated by the thermal radiation from the atmosphere. From 8 to 13 µm, the only incoming radiation is from the cold universe. Beyond 13 µm, the incoming radiation is again dominated by the thermal radiation from the atmosphere.

By enabling or disabling specific heat-exchange channels, one can maximize the radiative condensation capability of the condenser, which can be quantified by the condensation power density as

$$q_{cond} = \int d\Omega\ \cos\theta \int_0^\infty d\lambda\ \left(I_{BB}(T_{cond},\lambda) - I_{atm}(T_{amb},\lambda) - I_{AM1.5}(\lambda)\right)\epsilon_{cond}(\lambda,\theta) \quad (1)$$

Here $\epsilon_{cond}(\lambda,\theta)$, $I_{BB}(T,\lambda)$ and $I_{AM1.5}(\lambda)$ are the angle-dependent absorptivity/emissivity of the condenser, the spectral irradiance of a blackbody at temperature $T$, and the AM1.5 solar spectral irradiance, respectively. $I_{atm}(T_{amb},\lambda) = I_{BB}(T_{amb},\lambda)\epsilon_{atm}(\lambda,\theta)$ is the spectral irradiance of the atmosphere at ambient air temperature $T_{amb}$, where $\epsilon_{atm}(\lambda,\theta) = 1 - t(\lambda)^{1/\cos\theta}$ is the angle-dependent emissivity of the atmosphere[26], where $t(\lambda)$ is the atmospheric transmittance in the zenith direction[27]. Here we assume $T_{amb} = 20\ °C$ throughout our calculation[28].

A blackbody has all four channels open for radiative exchange, as shown in Fig. 1c. Thus, the radiative heat radiated by the blackbody reaches 1100 W m$^{-2}$ at 100 °C. However, the blackbody also receives all the heating power density in the four channels, which reaches 1230 W m$^{-2}$ because the absorption of solar radiation (1000 W m$^{-2}$) and atmospheric radiation. Thus, a blackbody has no daytime condensation power, even for blackbody at 100 °C (Fig. 1d). At the other extreme, Fan et al. designed radiative emitters that close all channels except the universe[28]. The radiative heat received by these radiative emitters is reduced to almost zero, allowing cooling to well below the ambient temperature[23,25]. However, because the daytime sub-ambient radiative emitter only emits in the spectral region from 8 to 13 µm, its cooling power is also substantially reduced compared to that of a blackbody, leading to limited condensation power (Fig. 1d). This design is sub-optimal for a radiative condenser, where the figure of merit is maximum condensation power rather than minimum achievable temperature. Our radiative condensers only close the solar channel and leave the atmospheric channels completely open (Fig. 1c). Though this design is not ideal for sub-ambient cooling[29], the vapor created in most solar water-purification systems is at or above the ambient temperature, and thus an open atmospheric channel can contribute substantially to condensation. As shown in Fig. 1d, radiative condensers (red line) more than double the cooling power when compared to sub-ambient radiative emitters (blue line) at most vapor temperatures of practical relevance, e.g., in solar water-purification systems.

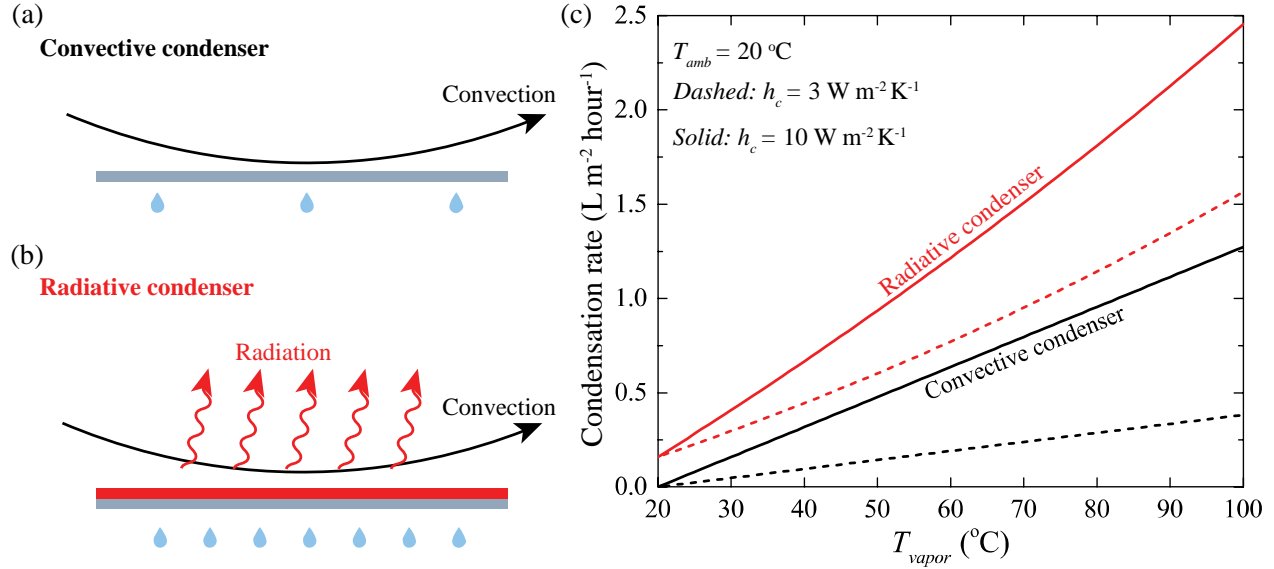

Figure 2. (a) and (b) Schematic of a convective condenser (a) and our radiative condenser (b). The convective condenser dissipates heat through only convection, while the radiative condenser dissipates heat through both convection and radiation. (c) Theoretically calculated condensation rates of the convective (black) and radiative condenser (red), assuming an ambient temperature of 20 °C and relative humidity of 100%.

Next, we discuss the application of our daytime radiative condenser for solar water-purification systems. The steady-state condensation rates are obtained from our model, which is available in Supplementary Note 1.

Water vapor created by solar absorption can condense through natural convection (Fig. 2a) because the vapor temperature $T_{vapor}$ in solar water-purification systems is above the ambient temperature $T_{amb}$. The convective cooling power density can be calculated as $P_{conv} \cong h_c(T_{vapor} - T_{amb})$, where $h_c$ is the convective heat-transfer coefficient. $h_c$ depends on the wind speed at the surface of the condenser[30], and usually ranges from 3 to 10 W m$^{-2}$ K$^{-1}$ for wind speed from 0 to 10 mph. The condensation rate $W_{water}$ can be calculated from the cooling power as $W_{water} = P_{conv}/\Delta_{vapor}$, where $\Delta_{vapor} = 2.26 \times 10^6$ J kg$^{-1}$ is the latent heat of vaporization. In most practical situations, the vapor temperature is well below 100 °C. For instance, the temperature of water vapor in the solar water-purification system in Ref. [8], which was demonstrated to have the highest solar-to-thermal efficiency, is only 40 °C. At such low temperatures, the upper bound of the condensation rate of a convective condenser is less than 0.1 L m$^{-2}$ hour$^{-1}$. This low condensation rate becomes the bottleneck of water production in solar water-purification systems. Even in the situation most favorable for convective cooling, e.g., vapor at 100 °C and wind speed of 10 mph (black solid curve in Fig. 2b), the condensation rate is only 1.3 L m$^{-2}$ hour$^{-1}$, which is still below the limit of the one-sun vapor-generation rate of 1.6 L m$^{-2}$ hour$^{-1}$ [11]. The situation becomes much worse when there is no wind (black dashed curve in Fig. 2b). Conversely, our daytime radiative condenser can substantially improve the condensation rate. Even at low temperatures, e.g. 40 °C., the condensation rate without wind is enhanced by more than 4 times, to 0.44 L m$^{-2}$ hour$^{-1}$. At high temperatures, e.g., 100 °C, the condensation rate almost doubles, reaching 2.5 L m$^{-2}$ hour$^{-1}$, well above the theoretical limit of the one-sun evaporation rate. Such a high condensation rate will also increase the vapor-pressure gradient inside a solar water-purification system, further facilitating the evaporation process.

We now describe our experimental realization of daytime radiative condensers. Figure 3a shows the schematic of a simple, large-area radiator designed to approach the near-ideal condenser absorptivity/emissivity spectrum (black dashed line). It consists of layers of polydimethylsiloxane (PDMS) and silver (Ag) on an aluminum (Al) substrate, with thickness of 100 μm, 150 nm, and 1 mm, respectively. The thermal radiation is primarily emitted by the PDMS layer, which has a near-unity emissivity for wavelengths longer than 4.5 μm due to Si–O and Si–C bond vibrations, given sufficient film thickness[31]. Simultaneously, PDMS is transparent to sunlight, which is efficiently reflected by the Ag layer. The Al substrate is chosen because of its high thermal conductivity and low price. In our experiment, the width and length of the condensation region are 25 cm and 20 cm, respectively. The spectral emissivity of the structure is characterized using Fourier transform infrared (FTIR) spectroscopy[32], shown in Fig. 3a. Our daytime radiative condenser reflects almost 96% of the solar radiation (0.3~4 μm) and emits efficiently in mid-IR region (>4 μm). We placed the radiator inside an insulating box made from polystyrene, as shown in Fig. 3b. The external surface of the insulating box is covered with aluminum tape to limit solar heating. A low-density polyethylene film covers the opening of the insulating box to reduce convective heat losses.

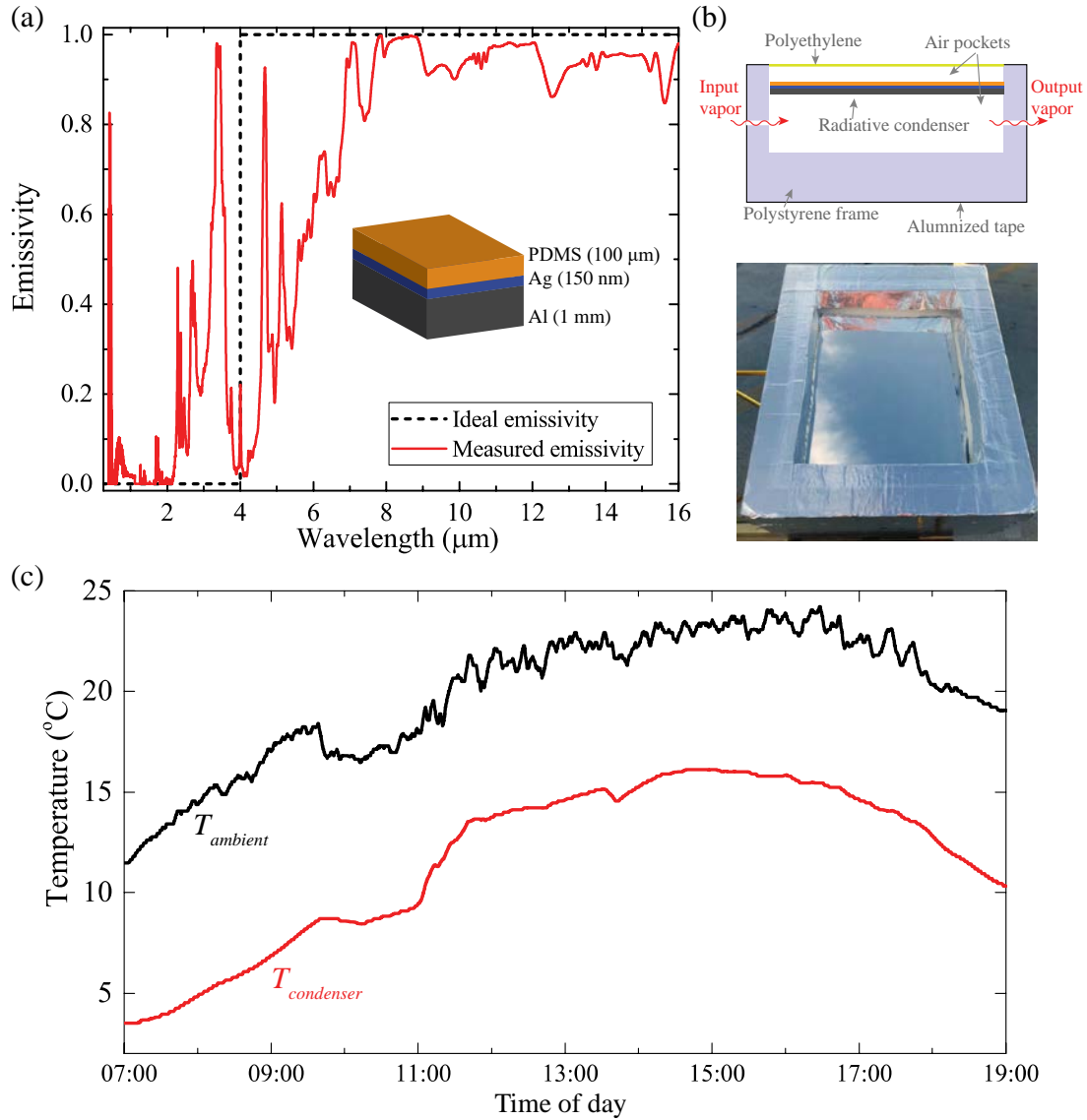

Figure 3. Experimental realization of our daytime radiative condenser. (a) Schematic of the daytime radiative emitter and the measured normal-incidence emissivity spectrum. The emitter consists of a 100-μm layer of PDMS, a 150-nm layer of silver, and a 1-mm-thick aluminum plate. (b) Experimental setup. The emitter is placed inside an insulating polystyrene box. The opening of the insulating box is covered by a thin polyethylene film and the external surface of the box is covered by aluminized-foil tape. (c) Measured temperature under direct sunlight, which is about 8 °C lower than the ambient temperature throughout the day. The measurement was performed on the roof of the Space Science and Engineering Building at the University of Wisconsin – Madison on September 29th, 2017.

To characterize the cooling effect of our radiative condenser under direct sunlight, we placed it on a roof facing the sky. The temperature of the condenser is measured by attaching a thermocouple at center of the backside of the condenser with conductive tape. The temperature of the ambient air is measured by placing a thermocouple inside a weather shield to avoid sunlight and wind. The measurement was performed on a sunny day with clear sky from 07:00 to 19:00. Figure 3c shows the temperature of the condenser (red curve) and the ambient air (black curve). Our condenser achieves a temperature reduction, i.e., the difference between the temperatures of the condenser and the environment, of about 8 °C throughout the day, which is slightly lower than that of existing sub-ambient radiative emitters[33].

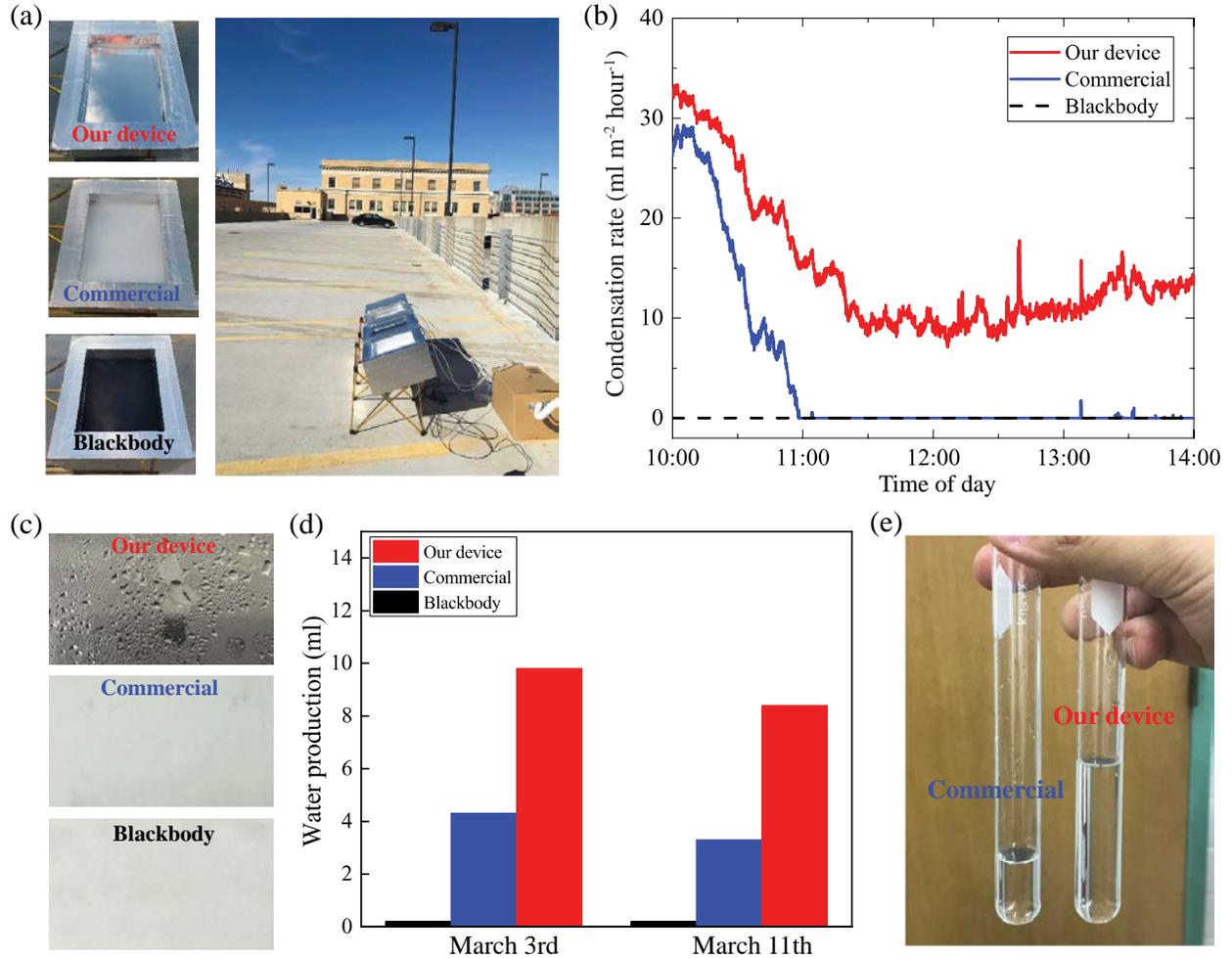

Figure 4. (a) Left: front-side photographs of condensers used in our measurements. A blackbody and a commercial radiative dew condenser (OPUR foil) are used for comparison. Right: Outdoor experimental setup. The condensers are placed on the roof of a parking ramp at the University of Wisconsin – Madison, under direct sunlight. (b) Real-time condensation rates of our daytime radiative condenser (red line), the commercial radiative condenser (blue line) and a blackbody (black line). The measurement was performed on March 10[th]. The blackbody had zero condensation rate due to absorption of sunlight. The commercial condenser initially had a non-zero condensation rate in the morning, but dropped to zero around 11:00. Conversely, our daytime condenser remained functional throughout the day. (c) Photos of the condensing surface for each condenser taken around 17:00 on March 11[th]. (d) Daily water production measured on two different days over 24-hour period. (e) Comparison between the amount of water condensed by our device and the commercial condenser.

Two additional condensers were used for comparison: a blackbody and a commercial radiative dew condenser [34] as shown in Fig. 4a. The blackbody is made by painting a thick layer of graphite-based carbon ink on top of an unpolished Al plate, which has the same dimension as our radiative condenser. The commercial condenser is based on a standard material for radiative dew condensation recommended by the International Organization for Dew Utilization (OPUR), known as OPUR foil[35,36]. The OPUR foil is a white low-density polyethylene foil, with 5% volume of TiO$_2$ nanoparticles (diameter 0.19 μm) and 5% volume of BaSO$_4$ nanoparticles (diameter 0.8 μm). We attached the OPUR foil to an Al plate with the same dimensions as our radiative condenser. In addition to those condensers, a reference device consisting of a plain Al plate was used to measure the ambient temperature and humidity of the input air. These condensers are placed inside insulating boxes with the same dimensions as our radiative condenser. For the reference device, the insulating box is completely covered by aluminum-foil tape to block all radiative heat-exchange channels.

All condensers were placed on a roof facing the sky. Humified air with a relative humidity of 90% ~ 95% was pumped into all condensers at a constant rate of $V_{in} = 0.9$ m$^3$ hour$^{-1}$. The vapor was filtered through a water trap to ensure no water droplets were contained in the vapor entering the cooling chamber. We performed day-to-night measurements on March 3$^{rd}$, 10$^{th}$ and 11$^{th}$, 2018. The temperature $T_{out}$ and relative humidity $H_{out}$ of the output airflow were measured by directly attaching temperature and relative-humidity probes at the outlet. The amount of water contained in the output air flow then was obtained as $m_{out} = H_{out} P(T_{out}) V_{in} M_{water}/R$, where $P(T)$ is the vapor pressure at temperature $T$, $R$ is the ideal gas constant, and $M_{water}$ is the molar mass of water. The vapor pressure $P(T)$ was calculated using the Buck equation[37]. The condensation rate then was obtained as $W_{cond} = (m_{in} - m_{out})/A_{cond}$, where $m_{in}$ is the amount of water contained in the input air flow and $A_{cond}$ is the area of the condensers. Due to the small area of our condensers (0.05 m$^2$), we did not extract the water that was produced. Instead, we obtained the overall production by measuring the weight change of the condenser.

Figure 4b shows a typical measurement during daytime. The blackbody absorbs almost all the sunlight and does not condense vapor (black curve). The commercial condenser absorbs less sunlight, and is able to condense vapor in the morning under relatively weak sunlight. Around 11:00, the condensation rate of the commercial condenser (blue curve) also dropped to zero. In contrast, our daytime condenser continued to condense water vapor throughout the day. As shown in Fig. 4d, the daily water production of our daytime radiative condenser is almost twice that of the commercial condenser. On the other hand, the total water production of the blackbody condenser was almost zero because the water condensed at night evaporated during the day.

A radiative condenser can also be designed to be transparent to solar radiation, and still be highly emissive in the infrared. Such a radiative condenser provides similar condensation performance as the device in Fig. 3 and 4, and can be directly integrated with existing solar stills. An example of transparent radiative condenser is given in Supplementary Note 2.

In conclusion, we demonstrate a passive device—a daytime radiative condenser—that can significantly accelerate the condensation of water vapor. We experimentally demonstrated water condensation of ambient-temperature vapor under direct sunlight, which cannot be realized by either conventional radiative dew condensers or convective condensers. Daytime radiative condensers can be incorporated into solar water-purification technologies, increasing total water production by more than a factor of two over the state of the art. Such technology is critically needed in areas where the sun is plentiful but clean drinking water is scarce.

Acknowledgements: M. K. acknowledges support from the National Science Foundation (ECCS-1750341). Z. Y. and M. Z. acknowledges support from the National Science Foundation (CMMI-156197). M. Z. acknowledges support from the 3M fellowship.

# Supplementary Information

**Supplementary Note 1. Steady-state model of passive condensation systems**

We developed a steady-state model to calculate the condensation rate of passive condensation systems. Below we describe it in detail.

As we discussed in the manuscript, the major cooling sources in passive condensers that operate without additional energy input are convection and radiation. The cooling power from conduction is generally small comparing to that from convection and is not considered here. The total cooling power density $q_{cooling}$ is given by

$$q_{cooling}(T_{cond}) = h_c(T_{cond} - T_{amb}) + \int d\Omega \cos\theta \int_0^\infty d\lambda\, I_{BB}(T_{cond}, \lambda)\epsilon_{cond}(\lambda, \theta) \quad (1)$$

Here $h_c$ is the convective heat transfer coefficient, $T_{cond}$ and $T_{amb}$ are the temperature of the condenser and the surrounding environment, respectively. $I_{BB}(T, \lambda)$ is the spectral intensity of a blackbody at temperature $T$. $\epsilon_{cond}(\lambda, \theta)$ is the angle-dependent spectral absorptivity/emissivity of the condenser.

The major heating sources are the solar radiation and atmospheric radiation. The total cooling power density $q_{heating}$ is given by

$$q_{heating} = \int d\Omega \cos\theta \int_0^\infty d\lambda \left(I_{AM1.5}(\lambda) + I_{BB}(T_{amb}, \lambda)\epsilon_{cond}(\lambda, \theta)\right)\epsilon_{cond}(\lambda, \theta) \quad (2)$$

Here $I_{AM1.5}(\lambda)$ is the AM1.5 solar spectral irradiance and $T_{amb}$ is the ambient air temperature. The angle-dependent emissivity of the atmosphere is given by[1] $\epsilon_{atm}(\lambda, \theta) = 1 - t(\lambda)^{1/\cos\theta}$, where $t(\lambda)$ is the atmospheric transmittance in the zenith direction[2].

As humidified air with temperature $T_{vapor}$ and relative humidity $H_{relative}$ flows through the condenser, it is cooled down. The vapor pressure decreases as the temperature decreases. Condensation occurs when the vapor pressure reaches the saturation vapor pressure. We approximated the saturation vapor pressure $P(T)$ at temperature $T$ using the Buck equation[3]:

$$P(T) = 611.21 \exp\left(\left(18.678 - \frac{T - 273.15}{234.5}\right)\left(\frac{T - 273.15}{T - 16.01}\right)\right) \quad (3)$$

The amount of power density required for condensation vapor condensation is given by

$$q_{vapor}(T_{cond}) = C_{V,air}u(T_{cond} - T_{vapor}) + u\Delta_{vap}\left(\frac{H_{relative}P(T_{vapor})}{RT_{vapor}} - \frac{P(T_{cond})}{RT_{cond}}\right) \quad (4)$$

Here $C_{V,air}$ is the specific heat capacity of air at constant volume $u$ is the speed of the input air flow at the air-condenser interface. $R$ is the ideal gas constant. The latent heat from vapor to liquid water is given by $\Delta_{vap}$= 40.63 kJ/mol.

At steady state, the whole system reaches thermal equilibrium, which satisfies

$$q_{cooling}(T_{cond}) - q_{heating} = q_{vapor}(T_{cond}) \tag{5}$$

By solving Eq. 5, we obtain the working temperature of the condenser $T_{cond}$, which depends on the input humidity $H_{relative}$ and air flow rate $u$. The condensation rate of the condenser then can be calculated as

$$W_{water} = \frac{q_{vapor}}{\Delta_{vap}} M_{water} \tag{6}$$

where $M_{water}$ is the molar mass of water.

To validate our theoretical model, we predict the condensation rate of our condenser based on measured $T_{in}$, $T_{amb}$ and $H_{in}$, and compare it to the measurement. The measurement was performed from March 10$^{th}$ to 11$^{th}$ on the roof a parking ramp of University of Wisconsin – Madison. The input air flow rate $u = 0.025$ m/s. The convective heat transfer coefficient $h_{conv}$ is taken to be $6\ \text{Wm}^{-2}\text{K}^{-1}$ to fit the experimental data. The results are plotted in Fig. S1. The predicted condensation rate (red curve) fits the measurement (black curve) very well. Here for simplicity, we consider nighttime condensation where the solar radiation is completely suppressed.

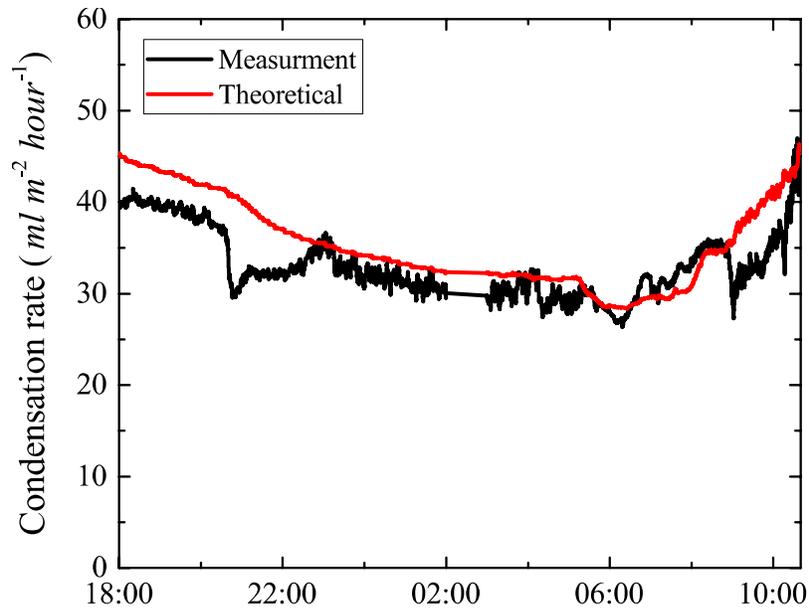

Figure S1. The theoretically predicated nighttime condensation rate (red) fits the experimental measurements (black) very well.

**Supplementary Note 2. Transparent daytime radiative condenser**

Here we propose a simple design of a transparent daytime radiative condenser that can be readily implemented in existing solar stills. As shown in Fig. S2a, the transparent radiative condenser consists of a thin layer of PDMS, which has a thickness of 100 µm, on top of glass. Figure S2b shows the transmission (red curve) and emissivity (black curve) spectra of the PDMS-glass condenser, which is more than 93% transparent to solar radiation. The structure also has near-unity emissivity in the mid-IR region. As a result, it has similar radiative condensing performance as the daytime radiative condenser described in Fig. 4 in the manuscript.

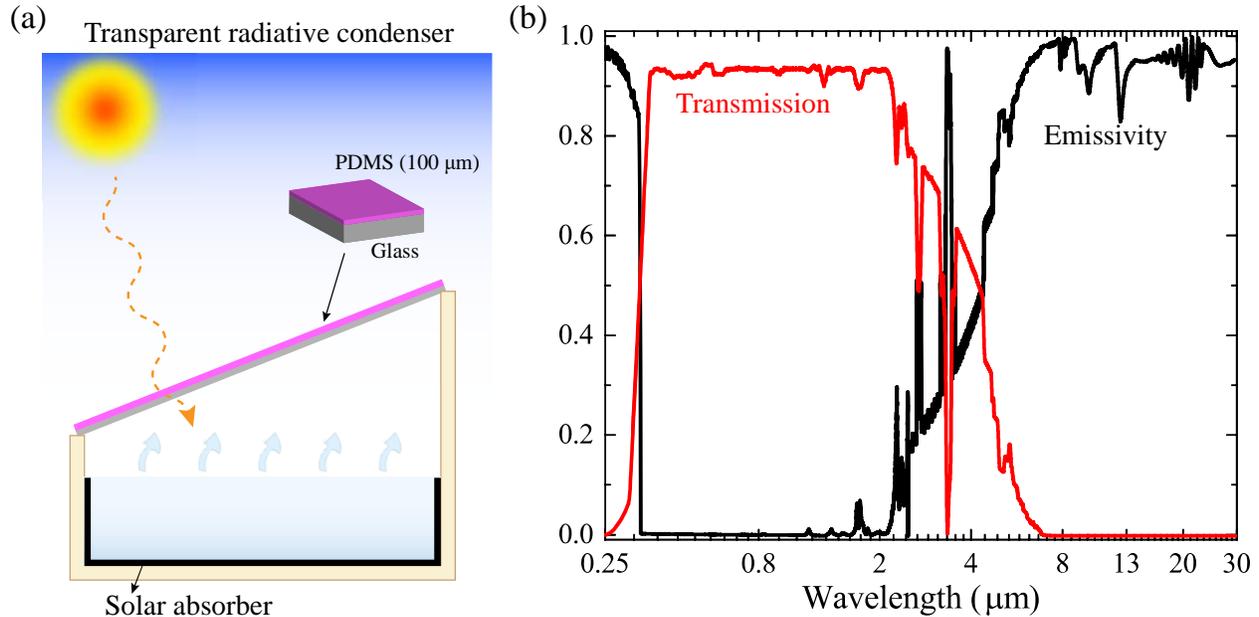

Figure S2. Transparent radiative condenser. (a) The transparent condenser can be readily implemented in existing solar stills. The structure is a layer of PDMS with a thickness of 100 µm on top of a glass substrate. (b) Transmission (red) and emissivity (black) spectra.